\documentclass[11pt,a4paper,twocolumn]{article}
\usepackage[utf8]{inputenc}
\usepackage{amsmath}
\usepackage{amsfonts}
\usepackage{amssymb}
\usepackage{makeidx}
\usepackage{graphicx}
\usepackage{lmodern}
\usepackage{textcomp}
\usepackage{gensymb}
\usepackage{siunitx}
\usepackage[version=4]{mhchem}
\usepackage[noblocks]{authblk}
\usepackage{makecell}
\usepackage{hhline}
\usepackage{changepage}
\usepackage{csquotes}
\usepackage[margin=1.5cm, bottom=2cm]{geometry}
\usepackage{bookmark}

\usepackage[doi=false,isbn=false,
		url=false,eprint=false,
		style=authoryear-icomp,
		citestyle=authoryear,backend=biber,
		natbib=true,
		maxcitenames=2,
		uniquename=false,
		uniquelist=minyear]{biblatex}
\DeclareFieldFormat[article,periodical]{volume}{\mkbibbold{#1}}

\renewbibmacro{in:}{}	

\newcommand*{\mkandothers}{\mkbibemph}
\renewbibmacro*{name:andothers}{%
  \ifboolexpr{
    test {\ifnumequal{\value{listcount}}{\value{liststop}}}
    and
    test \ifmorenames
  }
    {\ifnumgreater{\value{liststop}}{1}
       {\finalandcomma}
       {}%
     \andothersdelim\bibstring[\mkandothers]{andothers}}
    {}}
    
\DeclareFieldFormat[article]{title}{}

\AtEveryBibitem{%
  \clearfield{number}}
  
\renewcommand{\vec}[1]{\mathbf{#1}}

\title{Quantitative powder diffraction using a (2+3) surface diffractometer and an area detector}
\date{}
\author[1,2,*]{Giuseppe Abbondanza}
\author[1]{Alfred Larsson}
\author[4]{Francesco Carlá}
\author[1]{Edvin Lundgren}
\author[3,*]{Gary S. Harlow}
\affil[1]{Division of Synchrotron Radiation Research, Lund University, 221 00 Lund, Sweden}
\affil[2]{NanoLund, Lund University, 211 00 Lund, Sweden}
\affil[3]{Department of Chemistry, University of Copenhagen, 2100 Copenhagen, Denmark}
\affil[4]{Diamond Light Source, Didcot OX11 0DE, United Kingdom}
\affil[*]{Corresponding authors: giuseppe.abbondanza@sljus.lu.se, gary.harlow@chem.ku.dk}

\begin{document}

\maketitle

\begin{abstract}
X-ray diffractometers primarily designed for surface x-ray diffraction are often used to measure the diffraction from powders, textured materials, and fiber-texture samples in so-called $2\theta$ scans. Unlike high-energy powder diffraction only a fraction of the powder rings is typically measured and the data consists of many detector images across the $2\theta$ range.  Such diffractometers typically scan in directions not possible on a conventional lab-diffractometer, which gives enhanced control of the scattering vector relative to the sample orientation. There are, however, very few examples where the measured intensity is directly used, such as for profile/Rietveld refinement, as is common with other powder diffraction data. Although the underlying physics is known, converting the data is time-consuming and the appropriate corrections are dispersed across several publications, often not with powder diffraction in mind. In this paper we present the angle calculations and correction factors required to calculate meaningful intensities for $2\theta$ scans with a (2+3)-type diffractometer and an area detector. We also discuss some of the limitations with respect to texture, refraction, and instrumental resolution, and what kind of information one can hope to obtain.
\end{abstract}

\section{Introduction}		\label{sec:intro}

\begin{figure}[ht!]
\includegraphics{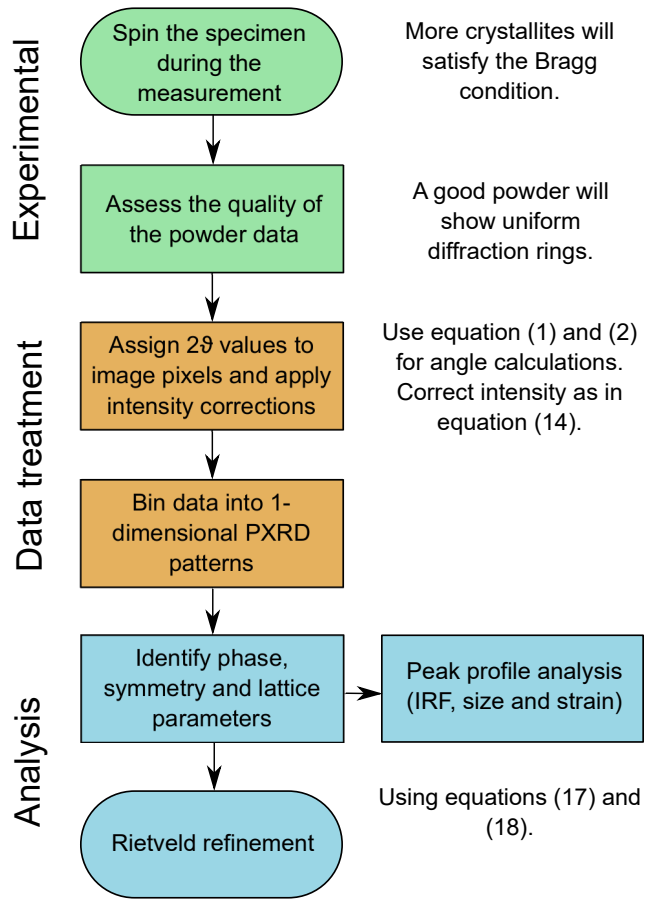}
\caption{Flowchart outlining the steps and the relevant equations involved in the investigation of powdered samples from the experiment to the determination of the structure.}
\label{fig:flowchart}
\end{figure}

\par Since the early studies by \citet{debye1916}, powder x-ray diffraction (PXRD) has become a well-established characterization technique. It has proved to be a fundamental tool for phase identification and structure determination of materials. Quantitative analyses of PXRD data enables access to information such as size, strain and stress of the crystallites, the number of different phases in multi-phases materials, atomic and unit lattice parameters. PXRD data quality improves significantly when synchrotron x-ray beams are employed, which provide: a high photon flux, enhanced collimation, tunable energies, and a superior angular resolution. In surface x-ray diffraction (SXRD), an experimental setup composed of a (2+3) diffractometer, an area detector is used at some beamlines. The (2+3) diffractometer was  presented by \citet{vlieg1997} and its combination with an area detector was explored by \citet{schleputz2005}. Although this setup was originally designed for single-crystal SXRD, it can also be used in PXRD and in grazing-incidence x-ray diffraction (GIXRD) by rotating the detector about the diffractometer center in longitudinal and equatorial $2\theta$ scans, across the Debye-Scherrer cones. In this scenario, the major benefit of the (2+3) diffractometer is control over the orientation of the sample and of the detector (and thus of the scattering vector) which enables convenient investigations of specimen textures and preferred orientations. This experimental setup is thus well-suited for a wide range of sample types with an extended face, such as thin polycrystalline films on substrates. The control over the grazing-incidence angle and thus over the x-ray penetration depth enables the study of layered materials and buried interfaces. Furthermore, the relatively high resolution of the instrument (compared to bench-top diffractometers) enhances the study and the identification of multi-phase systems.

\par The phenomena involved in PXRD are well-known and have been extensively studied in the last century. A significant step forward in the analysis of PXRD data was the Rietveld method \citep{rietveld1966, rietveld1967, young1993, vanlaar2018}. Such refinement is usually performed on PXRD patterns where the intensity is plotted as a function of $2\theta$. Therefore, with the (2+3) setup at the I07 beamline \cite{nicklin2016}, it is necessary to integrate the two-dimensional data collected by the area detector into a one-dimensional pattern. Furthermore, a series of intensity corrections should be applied to the measured  intensities, to obtain the structure factors that depend on the underlying crystallography of the sample. 

\par Area detectors with fixed position have been used extensively in powder diffraction, e.g., in the mapping of grain boundaries \citep{poulsen2001} and in texture analysis \citep{wenk2017}. Furthermore, software meant for the analysis of 2-dimensional diffraction data has been developed. For instance, Fit2D \citep{hammersley2016} and pyFAI \citep{kieffer2013} enable the integration of two-dimensional data to a one-dimensional pattern only for detectors with fixed position. The software BINoculars \citep{roobol2015, drnec2014} is tailored toward the analysis of surface x-ray diffraction data and is used to assign the intensity detected by every single pixel of an area detector to a bin in the three-dimensional reciprocal space. This means that it is possible to reduce the powder diffraction data collected by a (2+3) diffractometer with this software although it is computationally expensive and requires a backend.
\par In this work, we present the calculations and the correction factors  needed to extract quantitative information from $2\theta$ scans with (2+3) diffractometers and area detectors. The calculations are part of a process intended for the characterization of crystalline materials, as illustrated by the flowchart in Fig. \ref{fig:flowchart}. We assessed the validity of the calculations and of the correction factors by refining a \ce{LaB6} reference sample. Furthermore, we calculated the instrumental resolution function (IRF) of the setup and compared the integrated intensities collected at different experimental geometries, namely capillary in transmission, grazing-incidence and Bragg-Brentano.

\section{Experimental setup} \label{sec:exp_setup}

\par The experimental work was conducted at the beamline I07, a hard x-ray (8-30 keV) high resolution diffraction beamline at Diamond Light Source, UK \citep{nicklin2016}. The x-ray beam had an energy of 20 keV and a size of 100 $\mu$m vertically and 200 $\mu$m horizontally. To record the powder diffraction intensities, a Huber (2+3) diffractometer \citep{vlieg1998} and a Pilatus 100K detector were used.

\begin{figure}[ht!]
\caption{Schematic of a (2+3) diffractometer. The arrows point towards the positive direction of rotation in the case of horizontal (blue) or vertical (red) geometry and for the detector (black).}
\includegraphics{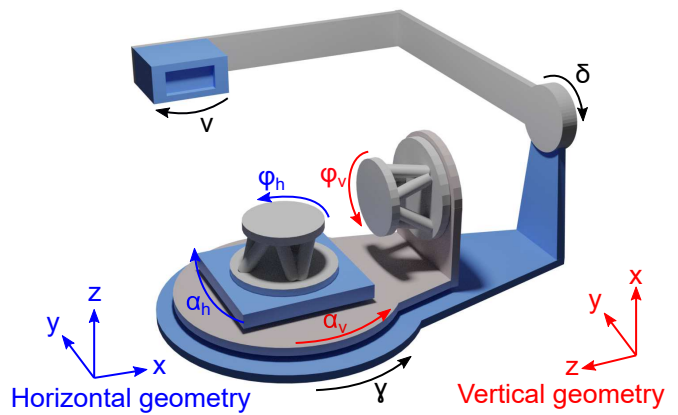}
\label{fig:diff}
\end{figure}

\par A schematic diagram of the Huber (2+3) diffractometer is shown in Fig. \ref{fig:diff}. The coordinate frame of reference depends on whether the sample geometry is horizontal (blue) or vertical (red). In both modes of operation the sample is mounted on a hexapod (Micos), which allows scanning of the sample translations and rotations for the initial alignment. The grazing-incidence angle of the synchrotron beam onto the sample surface and the azimuth are given by the rotation of $\alpha_h$ and $\phi_h$ in the horizontal geometry, and by $\alpha_v$ and $\phi_v$ in the vertical geometry.
\par The three detector circles are fully independent of the two sample circles and allow for radial scans in the horizontal plane ($\gamma$) or out of the horizontal plane ($\delta$), and rotation of the detector around its surface normal ($\nu$). 
\par The Pilatus 100K area detector consists of an array module of 487 $\times$ 195 pixels with size 172 $\mu$m $\times$ 172 $\mu$m, resulting in an active area of 83.8 $\times$ 33.5 mm and it can detect photons in an energy range of 3-30 keV with a dynamic range of $2^{20}$ \citep{kraft2009}.
\par Due to the size of the detector and to the energy range available in this experimental setup, the detection of whole Debye-Scherrer rings in image is not feasible. Therefore, the detector arm is usually scanned across the diffraction cones and small fractions of the Debye-Scherrer rings are detected.

\section{Angle calculations}	\label{sec:transformations}

\begin{figure}
\includegraphics{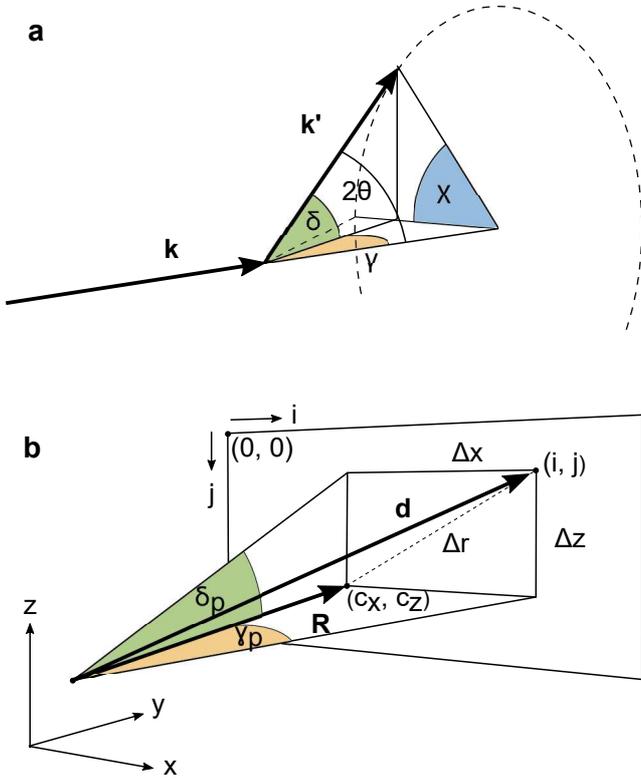}[ht!]
\caption{(a) Schematic representation of the angles $\gamma$, $\delta$, $2\theta$ and $\chi$, subtended by the incoming beam wavevector \textbf{k} and a generic scattering wavevector \textbf{k'} and (b) diagram showing the offsets $\Delta x$ and $\Delta z$ from the detector center of a generic $(i,j)$ pixel. The vector \textbf{d} (i.e., the distance of the pixel from the diffractometer center) subtends the angles $\delta_p$ and $\gamma_p$.}
\label{fig:vectors}
\end{figure}

\par With the setup described above, the Debye-Scherrer rings are detected by scanning the detector arm along $\gamma$ and $\delta$. The sequence of detector images collected during a scan can be integrated and processed to obtain diffracted intensities as a function of the diffraction angle $2\theta$ and the azimuthal angle $\chi$, illustrated in Fig. \ref{fig:vectors} (a). A detailed derivation of these angles for every pixel of an area detector are given in Appendix. According to this,  $2\theta$ and $\chi$ are expressed as

\begin{equation}
\label{eqn:2th_1}
2\theta = \arccos(\cos{\delta} \cos{\gamma}),
\end{equation}

\begin{equation}
\chi = \arctan\left(\frac{\tan{\delta}}{\tan{\gamma}}\right).
\label{eqn:chi_1}
\end{equation}

Fig. \ref{fig:vectors} (b) shows the angles $\gamma_p$ and $\delta_p$ subtended by a generic detector pixel $(i,j)$, located at a certain distance $d$ from the diffractometer center. Once the coordinates $(x_p, y_p, z_p)$ of this pixel are determined using the rotation matrices shown in Appendix, it is possible to calculate $\gamma_p$ and $\delta_p$ as follows:

\begin{equation}
\gamma_p = \arctan\left(\frac{z_p}{y_p}\right),
\label{eqn:gamma_p_1}
\end{equation}

\begin{equation}
\delta_p = \arcsin\left(\frac{x_p}{d}\right).
\label{eqn:delta_p_1}
\end{equation}

\begin{figure}[ht!]
\includegraphics{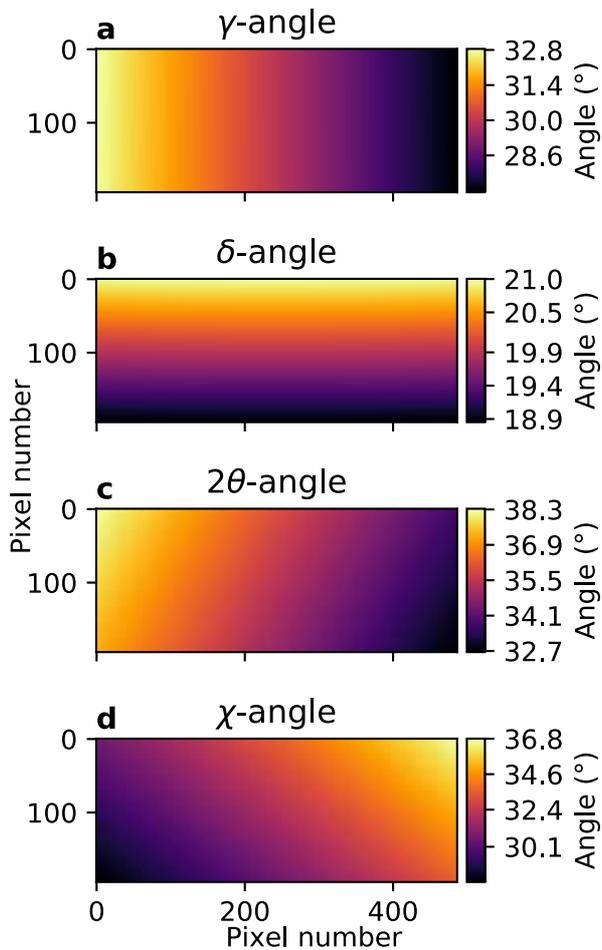}
\caption{2D plots of the angle values assigned to each pixel of a Pilatus 100K detector positioned at $\gamma$=30°, $\delta$=20° and R=897 mm. The plots show $\gamma_p$ (a), $\delta_p$ (b), $2\theta$ (c) and azimuthal $\chi_p$ (d).}
\label{fig:maps}
\end{figure}

Therefore, equations (\ref{eqn:gamma_p_1}) and (\ref{eqn:delta_p_1}) can be substituted into equations (\ref{eqn:2th_1}) and (\ref{eqn:chi_1}) to produce some angle maps like those shown in Fig. \ref{fig:maps}, calculated for the case where $\gamma$=30° and $\delta$=20°, assuming a detector distance of 897 mm, which is the distance at which the data were collected in this work.

\section{Intensity correction factors}	\label{sec:corr}

\par The intensity of a powder diffraction pattern can be expressed as

\begin{equation}
\label{eqn:int2theta}
I(2\theta) \propto \Phi_0 M_{hkl}\mid F_{hkl}(2\theta)\mid ^2P(\gamma, \delta) L(2\theta) V
\end{equation}

where $\Phi_0$ is the incident photon flux, $M_{hkl}$ is the multiplicity (i.e., the number of symmetry-equivalent reflections contributing to a single peak), $F_{hkl}$ is the structure factor, $P(2\theta)$ is the polarization factor, $L(2\theta)$ is the Lorentz factor and $V$ is the sample volume from which the diffracted intensity arises.
Deviations from this are often due to the presence of a preferred orientation or crystal texture.

\subsection{Polarization factor}	\label{sec:pol}

\par The x-ray beam produced by the I07 undulator has a strong horizontal polarization \citep{nicklin2016} and therefore the scans in the horizontal plane (i.e., $\gamma$-scans) are more affected by this factor than the scans out of the horizontal plane (i.e., $\delta$-scans). For this reason it is more convenient to express the polarization factor P as a function of ($\gamma$, $\delta$) using the following expression that has been presented by several authors \citep{vlieg1997,schleputz2005}

\begin{equation}
P(\gamma, \delta) = p_h(1 - \cos^2{\delta}\sin^2{\gamma}) + (1 - p_h)(1 - \sin^2{\delta})
\label{eqn:pol}
\end{equation}

The polar plot in Fig. \ref{fig:corr} (a) illustrates the variation of the polarization factor during an in-plane ($\gamma$) scan. Conversely, it will be less significant for an out-of plane ($\delta$) scan, where the polarization factor will be fairly constant.
\par In high energy x-ray diffraction, where the beam energy is usually in the 60-80 keV range and the whole Debye-Scherrer rings are detected, this factor is often neglected. With high energies, although the beam polarization is almost entirely horizontal (e.g., 98\%), the angular-dependent decay does not have a significant impact. Assuming a detector distance of 1.5 m and that a large flat panel detector is used (e.g., 40 $\times$ 40 cm), the highest $\vec{q}$ accessible has an absolute value of approximately 5.8 Å$^{-1}$. If the beam energy is 80 keV, the corresponding $2\theta$ angle will be 8.2° and the highest drop in intensity due to the polarization on the detector horizon can be calculated using equation (\ref{eqn:pol}) and equals 4\%. This small drop in intensity becomes even less significant when the intensity is integrated over the whole Debye-Scherrer ring. When lower energies are used (e.g., 20 keV), the $2\theta$ angle corresponding to a scattering vector of 5.8 Å$^{-1}$ is 33.2°. When the detector arm of a (2+3) diffractometer is scanned horizontally (i.e., $\delta$ = 0°) and reaches a $\gamma$ value of 33.2° , the polarization has dropped by 31\%. This decay must be taken into account especially since only a fraction of the Debye-Scherrer rings are detected.

\subsection{Lorentz factor}	\label{sec:lorentz}

\begin{figure}[ht!]
\includegraphics{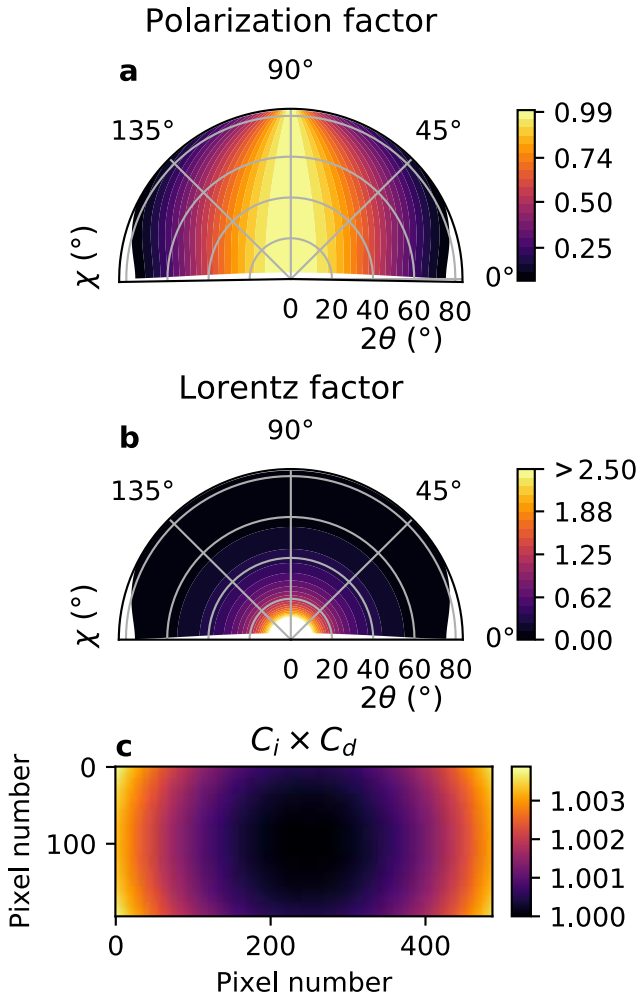}
\caption{Intensity correction factors calculated at a distance of 897 mm from the diffractometer center. Polar contours of polarization factor (a) and Lorentz factor (b), as a function of the diffraction and the azimuthal angles. Flat detector correction calculated for a Pilatus 100K with center pixel coordinates $(c_i,c_j) = (246,100)$.}
\label{fig:corr}
\end{figure}

\par A detailed derivation of the Lorentz factor was given by \citet{buerger1940}, where it is defined to be proportional to the time that a reflection stays in the Bragg condition. The Lorentz factor depends on the type of experiment performed and especially on the scanning variable employed to detect reciprocal space \citep{vlieg1997}. For instance, in single-crystal diffraction, the integrated intensity is measured by rotating the sample over the entire width of a reflection while the detector position is fixed, in a so-called rocking scan or $\Phi$-scan. In that case, the measured intensities need to be corrected by a geometrical factor, i.e. the Lorentz factor, that expresses the relative time spent by each point in reciprocal space in the reflecting position during the $\Phi$-scan. In PXRD, instead, the detector is rotated while the sample position is stationary, thus there is virtually no access to reciprocal space volume. However, due to the random orientation of the crystallites, a powder can be seen as a single crystal that is rotated along the $\Phi$-axis and an axis orthogonal to the $\Phi$ rotation.
\par In PXRD, the Lorentz factor L is the product of three terms. The first term is the \enquote{Darwinian} single-crystal part \citep{darwin1922}, that accounts for changes in integration volume as a function of $2\theta$:

\begin{equation}
L_1 = 1/\sin{2\theta}.
\label{eqn:L1}
\end{equation}

\begin{figure*}
\centering
\includegraphics[scale=0.9]{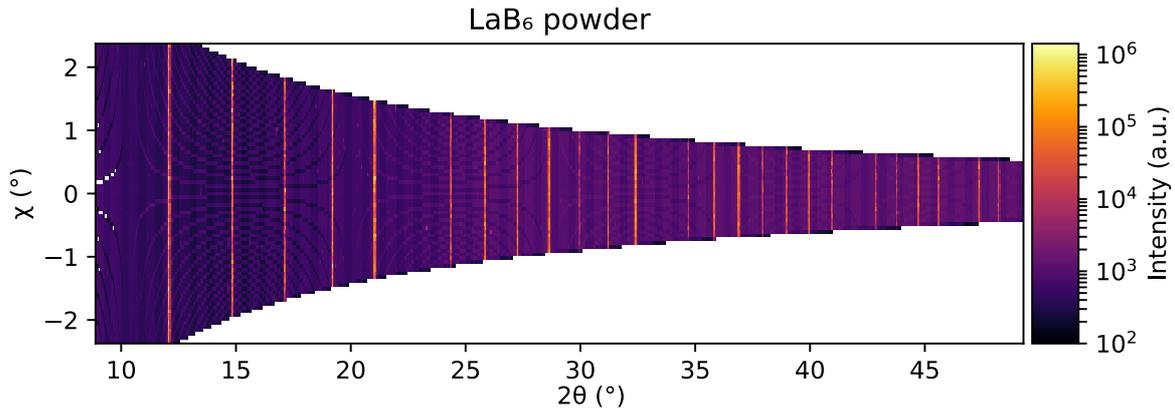}
\caption{Diffracted intensity as a function of $2\theta$ and $\chi$ originating from a NIST \ce{LaB6} SRM 660c encapsulated in a 1 mm diameter borosilicate capillary. The data was collected in the Debye-Scherrer geometry by scanning the detector in the horizontal plane ($\gamma$-scan).}
\label{fig:azi}
\end{figure*}

The second term is proportional to the fraction of diffraction ring detected for different $2\theta$ values. If we express the radius of the base of a generic Debye-Scherrer cone as $2\pi k \sin{2\theta}$, the fraction recorded by the detector is $k\Delta \chi/2\pi k \sin{2\theta}$ and therefore the second term will be proportional to

\begin{equation}
L_2 = \Delta \chi / \sin{2\theta},
\label{eqn:L2}
\end{equation}

where $\Delta \chi$ is the range of azimuthal $\chi$ values accessible by the detector at a given $\gamma$ and $\delta$ and it can be considered constant \citep{als2011}. This effect is evident in Fig. \ref{fig:azi}, where the diffraction intensity from a borosilicate capillary containing NIST \ce{LaB6} SRM 660c is plotted as a function of $2\theta$ and $\chi$. Here, the range of accessible $\chi$ values decreases significantly with the increase of $2\theta$.
\par The third term is proportional to the number of observable lattice points at the same time, and therefore to the circumference of the base-circle of the Debye-Scherrer cones. If we denote a particular reciprocal lattice vector with $\vec{G_{hkl}}$, this circumference is $G_{hkl}\sin{(\frac{\pi}{2} - \theta)}$=$G_{hkl}\cos{\theta}$. In the assumption that the crystallites and thus their reciprocal lattice points are homogeneously distributed on the Ewald sphere, this term is proportional to

\begin{equation}
L_3 = \cos{\theta}.
\label{eqn:L3}
\end{equation}

Note that $L_3$ is not proportional to the possible permutations of (h,k,l) since this is already accounted for by the multiplicity factor $M_{hkl}$ in equation (\ref{eqn:int2theta}).
The Lorentz factor is given by the product $L_1 L_2 L_3$ as in equation (\ref{eqn:L_before}), which is rearranged in equation (\ref{eqn:lorentz}):

\begin{equation}
\label{eqn:L_before}
L(2\theta) = \frac{1}{\sin{2\theta}} \frac{1}{\sin{2\theta}} \cos{\theta}
\end{equation}

\begin{equation}
\label{eqn:lorentz}
L(2\theta) =  \frac{1}{\sin{\theta}\sin{2\theta}} 
\end{equation}

As shown in Fig. \ref{fig:corr} (b), the Lorentz factor has a rather quick decay as a function of $2\theta$, regardless of the azimuthal angle $\chi$. It is perhaps the most significant intensity correction that needs to be applied to experimental data, especially at small $2\theta$.

\subsection{Flat detector corrections}	\label{sec:flat}

\begin{figure}[ht!]
\includegraphics{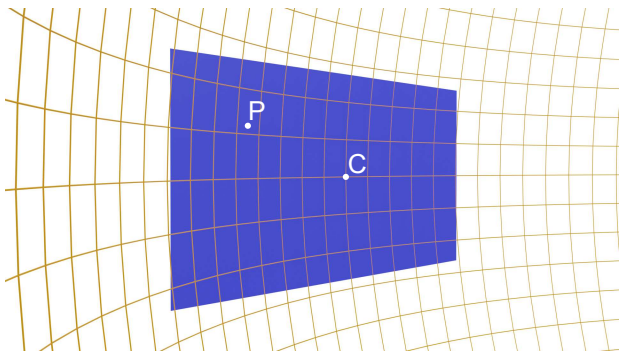}
\caption{The flat surface of the Pilatus 100K (blue rectangle) intersects a sphere centred in the diffractometer (orange frame) in the point C (the detector centre). A generic diffracted beam would impinge the point C normally and subtend a non-normal angle with any other generic point P.}
\label{fig:flat}
\end{figure}

In addition to the polarization and the Lorentz factor, the decrease of the subtended solid angle for the different pixels has to be taken into account. Two contributions normally describe this change. The first one is due to the fact that pixels away from the detector center are also further away from the diffractometer center, as depicted in Fig. \ref{fig:flat}. To account for this change in distance the measured intensities should be multiplied by 

\begin{equation}
C_d = \frac{d^2}{R^2}.
\label{eqn:Cd}
\end{equation}

$C_d$ is greater than unity for every pixel of the detector except the detector center C, where $d=R$. 
The second contribution is due to the non-normal incidence of the scattered beam owing to the flat detector surface and is given by \citet{schleputz2010}

\begin{equation}
\label{eqn:Ci}
C_i=\frac{1}{\cos{(\arctan{\Delta r /R})}},
\end{equation}

where $\Delta r = \sqrt{\Delta x^2 + \Delta z^2}$. 
\par For a better visualization of these correction factors, the product $C_d \times C_i$ is plotted in Fig. \ref{fig:corr} (c) for the same detector distance R. The correction becomes more significant close to the edges of the detector image, where it shows a maximum change of about 0.35 \% from the detector center. The significance of this correction increases as the detector distance decreases or as the detector size increases.

\subsection{Transmission corrections}	\label{sec:abs}
\par The sample volume that gives rise to diffracted intensity depends on the geometry of the experiment and is often a function of the scattering angle $2\theta$, the incidence angle $\alpha$ and the angle between the sample surface and the diffracted beam $\gamma = 2 \theta - \alpha$. The diffracting specimen volume is often modelled as an exponential aberration. Table 1 summarizes this correction for some common experimental geometries. As a general remark, the interaction volume is proportional to the area of illuminated sample, $A$, and is inversely proportional to the linear absorption coefficient, $\mu$. Furthermore, it should be noted that refraction leads to a reduction of the penetration depth and thus the interaction volume when the incidence angle is below the critical value. This effect is well-described by \citet{feidenhansl1989}, in a formalism that considers evanescent waves and their penetration into surfaces.

\begin{table*}
\centering
\caption{Volume corrections for some common experimental geometries.}
\centering
\begin{adjustwidth}{+0 cm}{-0.5cm}
	\begin{tabular}{ | c | c | c | c |}
	\hline
	Geometry	&	Equation	&	 Note 	&	References	\\
	\hhline{|=|=|=|=|}
	\makecell{Flat-plate \\ symmetric \\ (Bragg-Brentano)}	&	
	$V_{SR} = \frac{A}{2 \mu}$	&	-	&
	\makecell{\citep{cheary2004} \\ \citep{egami2003}}	\\
	\hline
	\makecell{Flat-plate \\ asymmetric \\ (grazing incidence)}	&
	$V_{AR} = \frac{A}{\mu} (1 + \frac{\sin\alpha}{\sin\gamma})^{-1}$	&	-	&
	\makecell{\citep{toraya1993} \\ \citep{james1967}}	\\
	\hline
	\makecell{Flat-plate \\ with finite \\ thickness}	&
	\makecell{$V = 2(1 + \frac{\sin\alpha}{\sin\gamma})^{-1}$ \\ $\{1 - \exp[-\mu t (\frac{1}{\sin\alpha} + \frac{1}{\sin\gamma}) ]\}$}	&
	\makecell{$t = $ thickness. \\ Symmetric geometry \\ when $\alpha = \gamma$}	&
	\makecell{\citep{egami2003}} \\
	\hline
	\makecell{Capillary in \\ transmission \\ (Debye-Scherrer)}	&
	\makecell{$A(\theta) = A_L \cos^2(\theta) + A_B \sin^2(\theta)$ \\
			$A_L = 2{I_0(z) - L_0(z) - \frac{I_1(z) - L_1(z)}{z}}$	\\
			$A_B = [I_1(2z) - L_1(2z)]/z$} &	\makecell{$z=2 \mu r$ \\ $I_{\nu}=$ $\nu$th-order \\
			modified Bessel \\ function	\\ $L_{\nu}=$ $\nu$th-order \\ modified Struve \\ function}	&
	\makecell{\citep{dwiggins1972} \\ \citep{sabine1998}}	\\ \hline
	\end{tabular}

\end{adjustwidth}

\label{tab:corr}
\end{table*}

\subsection{Summary of intensity corrections}	\label{sec:summary}
The corrected intensity is given by

\begin{equation}
\label{eqn:Icorr}
I_{corr} = I_{obs} \frac{C_d C_i}{LPV},
\end{equation}

where $I_{obs}$ is the observed intensity.

\section{Peak profile analysis}	\label{sec:width}

\par Analysis of the line-profile shape is useful for the determination of crystallite size and strain of a specimen. However, the width of a diffraction peak is a function of $2\theta$ and it depends not only on specimen properties but also on experimental conditions and sample size. Furthermore, instrumental features such as monochromators, collimating and refocusing optics, slits, beam divergence and energy bandwidth influence the broadening of diffraction peaks \citep{gozzo2006}. \par As the mentioned causes of broadening have either Gaussian or Lorentzian nature, the most used fitting function for PXRD peaks are based on Voigt or pseudo-Voigt line shapes. The full width at half maximum (FWHM) of the Voigt profile depends on the widths of the associated Gaussian and Lorentzian components $\Gamma_G$ and $\Gamma_L$ \citep{thompson1987,thompson1987_2}.

\subsection{Gaussian and Lorentzian broadening}	\label{sec:broadening}

\par The Gaussian widths contain information on the instrumental resolution function (IRF) and on the sample strain. An analytical description of the IRF was given by \citet{caglioti1958}:

\begin{equation}
\Gamma_r = {(U \tan^2\theta + V \tan\theta + W)}^{1/2}.
\label{eqn:caglioti}
\end{equation}

Specimen contributions to the Gaussian widths are the expression of crystal defects, dislocations and deformation of the unit cells, in what is known as inhomogeneous strain broadening:

\begin{equation}
\Gamma_{s} = 4 \epsilon \tan{\theta},
\label{eqn:strain}
\end{equation}

where the root mean square strain $\epsilon$ is a coefficient that depends on the elastic compliance and the mechanical properties of the specimen. Since this contribution is proportional to $\tan{\theta}$, some authors merged the strain contribution into equation (\ref{eqn:caglioti}). For example, \citet{thompson1987} enclosed the coefficient $\epsilon$ into the constant V, while \citet{wu1998} into the constant U. In this work we adopt the solution presented by \citet{thompson1987_2}, where the Gaussian width broadening is expressed as

\begin{equation}
\Gamma_G = (\Gamma_r^2 + \Gamma_s^2)^{1/2}.
\label{eqn:g_broad}
\end{equation}

The Lorentzian widths $\Gamma_L$ take into account the spectral bandwidth of the source and the sample crystallite size through the following equation \citep{cox1991}:

\begin{equation}
\Gamma_L = X \tan{\theta} + Y/ \cos{\theta}.
\label{eqn:l_broad}
\end{equation}

The X coefficient depends on the monochromating optics and it is in the order of magnitude of $10^{-4}$ for most of the synchrotron beamlines where Si(111) crystals are employed. The dependence of the bandwidth term on $\tan\theta$ can be derived by differentiating Bragg's law. 
The second term in equation (\ref{eqn:l_broad}) is the Scherrer crystallite size contribution. Here, $Y=K\lambda/D$ where $K$ is a dimensionless shape factor (generally close to unity), $\lambda$ is the wavelength of the x-ray radiation, and D is the crystallite size. One should remember that the use of equation (\ref{eqn:l_broad}) relates to the size of the coherent diffraction domains rather than the size of the crystallites \textit{per se} \citep{scherrer1912, patterson1939, hargreaves2016}.
\par Equation (\ref{eqn:l_broad}) can be rearranged as

\begin{equation}
\Gamma_L \cos\theta = X \sin\theta + Y.
\label{eqn:williamson-hall}
\end{equation}

\par Plotting $\Gamma_L \cos{\theta}$ against $\sin{\theta}$ in equation (\ref{eqn:williamson-hall}) would produce a line where the intercept depends on the crystallite size while the slope depends on the beam spectral bandwidth. Such a plot is known as a Williamson-Hall plot \citep{williamson1953} and an example of such analysis is given in \S \ref{sec:irf}.

\subsection{Beam footprint}	\label{sec:footprint}

\par
\begin{figure}[ht!]
\includegraphics[scale=0.8]{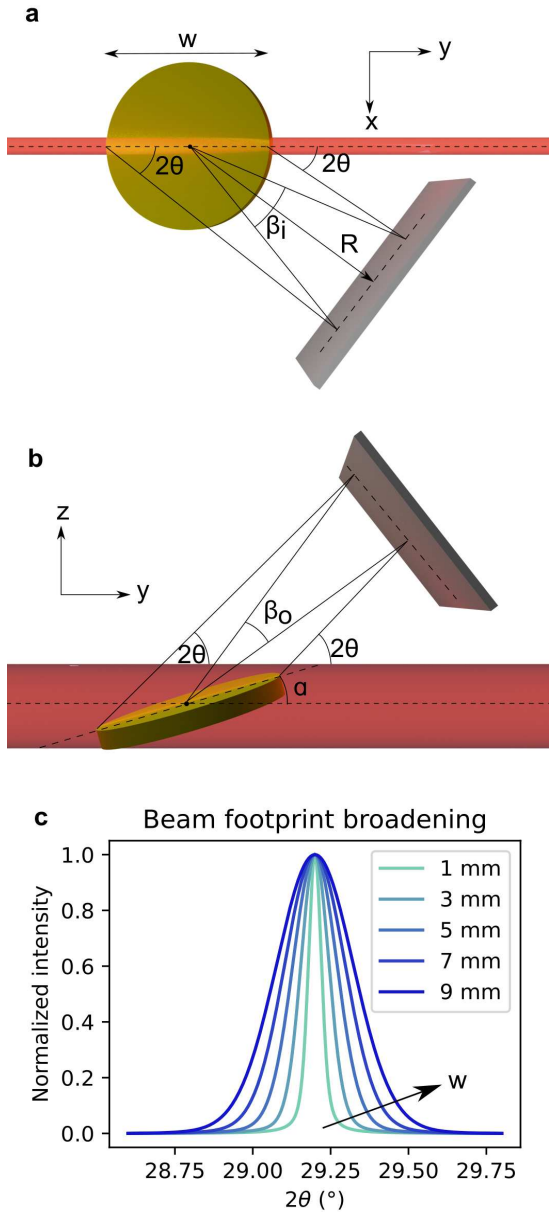}
\caption{Schematic depiction of the geometric broadening due to the beam footprint at grazing-incidence: top view of a $\gamma$-scan (a) and side view of a $\delta$-scan(b). Simulation of an Au(311) peak at 20 keV, at a grazing angle of 0.1°, scanned out-of-plane, for increasing sample widths (c).}
\label{fig:footprint}
\end{figure}

 The beam footprint on the sample causes a broadening that affects the instrumental resolution. This is especially true for grazing-incidence geometries, where the beam spills over the sample and illuminates it over its whole length. A depiction of this geometric effect is shown in Fig. \ref{fig:footprint}. Assuming that every volume element illuminated by the beam scatters, all intensity occurring at the diffraction angle $2\theta$ is spread out into a radial range $\beta$ on the detector. The angular spread for in-plane scans (Fig. \ref{fig:footprint} (a) and for out-of-plane scans (Fig. \ref{fig:footprint} (b) are respectively given by

\begin{equation}
\beta_i = 2 \arctan \left(\frac{w \sin 2\theta}{2 R}\right),
\label{eqn:footprint_a}
\end{equation}

\begin{equation}
\beta_o = 2 \arctan \left(\frac{w \sin (2\theta - \alpha)}{2 R}\right),
\label{eqn:footprint_b}
\end{equation}

where $w$ is the sample width and R is the sample-to-detector distance.
\par Fig. \ref{fig:footprint} (c) shows a simulated intensity profile of an Au(311) peak at 20 keV ($2\theta$ = 29.1996°), for different sample widths, assuming a scan out-of-plane and a grazing-incidence angle of 0.1°. The simulated peaks were calculated as Voigt profiles where the Lorentzian and Gaussian components were determined by the IRF of the I07 beamline at Diamond Light Source (see \S \ref{sec:irf}). The $\beta_o$ broadening was calculated using equation \ref{eqn:footprint_b} and summed to the Gaussian component of the Voigt profile as follows:

\begin{equation}
\Gamma_G = (\Gamma_r^2 + \beta_o^2)^{1/2}.
\end{equation}

\par A more insightful way to account for the beam footprint is to determine the peak profile change induced by specimen absorption. Equations (\ref{eqn:footprint_a}) and (\ref{eqn:footprint_b}) work on the assumption that the intensity of the incoming beam does not change significantly through the whole sample width $w$. This is not always true since the intensity decays exponentially as described by Beer-Lambert's law \citep{swinehart1962}. Therefore, a change in diffraction peak profile due to absorption can be modelled as an exponential function. Such effects are well-described for several diffraction geometries in the work of \citet{rowles2017}. Once the transmission profile function has been modelled, it can be convoluted with a Voigt or pseudo-Voigt profile in a refining algorithm. This approach not only works for correcting the beam footprint effect, but also accounts for possible peak asymmetries.
\par The beam footprint broadening can be limited by engaging the guard slits between the sample and the detector. If the aperture is small enough, the diffraction originating from a limited region of the sample is detected and the peak broadening is no longer angle-dependent, but it is rather defined by the slit aperture width. A description of the geometry involving detector slits is available in Appendix.

\subsection{Rietveld refinement}	\label{sec:rietveld}

\par Rietveld refinement (also known as \enquote{profile refinement}, \citep{vanlaar2018}) is a well-established analysis method for PXRD data and it is widely employed in the characterisation of polycrystalline materials \citep{rietveld1966, rietveld1967, young1993, loopstra1969, vanlaar1984}. This method consists of fitting the experimental data with a calculated intensity profile which is based on the structural parameters of the material. A non-linear least-squares algorithm (or other optimization strategy) finds the parameters for a theoretical profile that best matches the experimental intensities. The Rietveld method can be used to find unit cell parameters, phase quantities, crystallite size and strain, atomic coordinates and texture. Furthermore, it is possible to model texture and preferred orientations for example using spherical harmonics \citep{whitfield2009}, although this is outside the scope of the present article.
\par The quality of the data and having a good starting model are what mainly determines the success of the refinement. For a good PXRD measurement the x-ray beam size should be comparatively larger than the crystallite size. In this way, the statistical significance of the detected intensity is maximised due to a larger interaction volume, which leads to the formation of homogeneously continuous diffraction rings. In common PXRD setups the sample is spun to facilitate the measurement of uniform rings and this is also possible with (2+3)-type diffractometers by rotating the $\phi_h$ in the horizontal geometry or $\phi_v$ in the vertical geometry.
\par To perform Rietveld analysis, software like FullProf \citep{fullprof}, GSAS-II \citep{gsas} and DIFFRAC. SUITE TOPAS \citep{coelho2011} has been developed. In our somewhat unconventional case, where the data has been manually corrected by Lorentz and polarization factors, e.g. by using equation (\ref{eqn:Icorr}), it is possible to disable FullProf from applying further instrumental corrections by selecting \textit{Lorentz Polarization not performed} as the diffraction geometry. However, we could not determine how to disable the instrumental corrections in GSAS-II without modifying the Python code making such corrections. We did not test if this was possible in TOPAS. Another option would be to multiply the corrected data by the inverse Lorentz-polarization factors used in software such as GSAS and TOPAS. In this way, it should be possible to use these programs for Rietveld refinement of data collected on a (2+3) diffractometer.
\par In absence of preferred orientations, equation (\ref{eqn:int2theta}) describes the proportionality between the intensity of a diffraction pattern and the square modulus of the structure factor. Any deviations from this proportionality can be attributed to preferred orientations and refined in software like FullProf.
These deviations due to preferred orientations can be modelled by, e.g., the so-called modified March’s function \citep{dollase1986}. More complex models involve the generalized spherical-harmonic description \citep{sitepu2002}.
In order to find further evidence of preferred orientations, one could plot the intensity as a function of 2$\theta$ and $\chi$, as in Fig. \ref{fig:azi}. This complete view of all the data gathered in a detector scan enables the search for qualitative evidence of preferred orientations, such as intensity variations along the Debye-Scherrer rings. Based on the orientation of the scattering vector, it should be possible to quantify the preferred orientation.

\subsubsection{Refinement of a NIST \ce{LaB6} reference} \label{sec:riet_lab6}

\begin{figure*}
\centering
\includegraphics[scale=0.9]{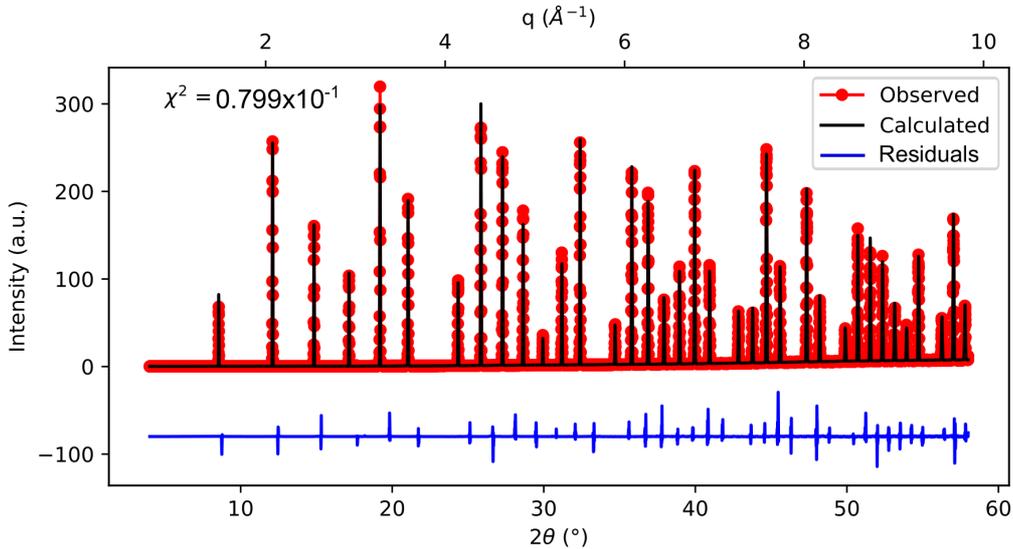}
\caption{Rietveld refinement of a NIST \ce{LaB6} SRM 660c PXRD pattern, measured with a 20 keV x-ray beam, performed with Fullprof using the pseudo-Voigt line shape. Before the refinement, the data was corrected by: Lorentz and polarization factors, flat-detector corrections and interaction volume of x-rays with a capillary. The background was subtracted by fitting a Chebychev polynomial (16 coefficients).}
\label{fig:refinement}
\end{figure*}

\par The \ce{LaB6} powder is a standard reference material commonly used in powder diffraction for the calibration of diffraction line positions and shapes. A sample of this powder was encapsulated in a borosilicate capillary of 0.5 mm radius, mounted vertically (i.e., with the capillary axis parallel to the z-axis) and measured in a series of $\gamma$ radial scans in a range of 5° to 60°. In order to measure continuous powder rings, the capillary was rotated by 0.5° along the z-axis (i.e., increasing $\phi_h$ in Fig. \ref{fig:diff}) between each scan. A total of 135 $\gamma$-scans were combined, each one at a different $\phi_h$.
\par The angle calculations presented in \S \ref{sec:transformations} were used to plot intensity against $2\theta$ in the 1-dimensional pattern shown in Fig. \ref{fig:refinement} and to correct the data by Lorentz-polarization factor, flat detector effects and interaction volume from a capillary in transmission geometry (see Table \ref{tab:corr}). Therefore the corrected data are proportional to the squared module of the structure factors and the multiplicity of each diffraction peak. Since the \ce{LaB6} standard has very well-defined unit cell parameters, the position of the diffraction peaks were used to calibrate the sample-to-detector distance and the nominal position of the $\gamma$ and $\delta$ motors.
\par The good agreement of the calculated model to the observed data is indicated by fairly low residuals (see Fig. \ref{fig:refinement}). This shows that the setup and data processing employed in this work can be used not only to investigate lattice parameters and phases, but also to record meaningful intensities proportional to structure factors and symmetry multiplicity. Deviations from this proportionality are a clear sign of texture and preferred orientation.
\par Rietveld refinement of a standard material thin film measured at grazing-incidence would be an interesting topic of discussion. However, the fabrication of such thin films often leads to sample morphology with prohibitive roughness. A film of \ce{LaB6} on a silicon wafer was produced in this work (the fabrication is described in § \ref{sec:refraction}). Due to the high roughness of this film, the control over the incidence angle, the interaction volume and the refraction effect was limited. The fact that Rietveld refinement was not feasible for this particular sample, does not exclude the possibility that it would be possible with a smooth polycrystalline film. In this case, it is recommended to select a grazing-incidence angle $\alpha$ above from the critical angle of the substrate, in order to avoid diffraction generated by the reflected beam, which would lead to two overlapping diffraction patterns with a shift of $2\alpha$.

\section{Further examples}	\label{sec:app}

\par The pattern in Fig. \ref{fig:refinement} was used to determine the IRF of the I07 beamline at Diamond Light Source. Furthermore, a different  \ce{LaB6} powder sample (Sigma-Aldrich, grain size ~10 $\mu$m), prepared by spin coating on a Si substrate, were measured in grazing-incidence and in Bragg-Brentano geometry.

\subsection{Instrumental resolution function of the I07 beamline}		\label{sec:irf}

\par As discussed in \S \ref{sec:broadening}, the peak widths contain instrumental as well as specimen-related information. Although software like FullProf and GSAS-II have built-in options to refine such parameters, a precise knowledge of the IRF is required to obtain significant information. In this work, we processed the data from the NIST \ce{LaB6} standard described in the previous section to calculate the IRF.
\par
\begin{figure}[ht!]
\includegraphics{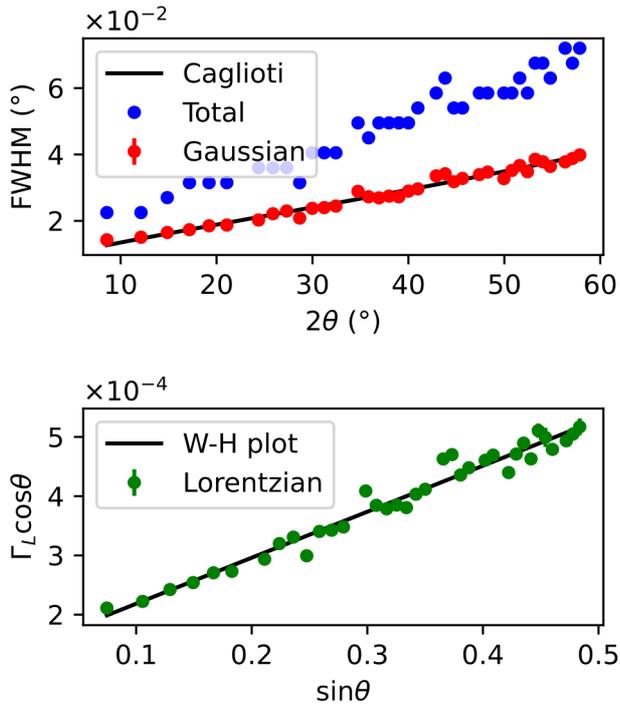}
\caption{Peak broadening analysis of a \ce{LaB6} standard reference material. The total FWHMs are plotted together with the Gaussian FWHMs, which are fitted by a Caglioti function [see equation (\ref{eqn:caglioti})] in (a), while the Lorentzian FWHMs, multiplied by $\cos \theta$, are fitted by a Williamson-Hall plot in (b).}
\label{fig:peak}
\end{figure}

 All the peaks in Fig. \ref{fig:refinement} were fitted one by one with a Voigt profile. The FWHM reported in Fig. \ref{fig:peak} (a) shows a broadening that has both Gaussian and Lorentzian contributions. The Gaussian and Lorentzian FWHMs of the peak widths were extracted and reported in Fig. \ref{fig:peak} in the form of a Caglioti function (a) and a Williamson-Hall plot (b). The dominance of Gaussian component in the broadening can be explained by the large average crystallite size i.e., above 1 $\mu$m as certified by NIST. Due to such a large crystallite size, the Scherrer contribution to the Lorentzian broadening is negligible and outside the limits imposed by coherent scattering domains \citep{miranda2018}. Furthermore, since the unit cell of \ce{LaB6} is not known to show inhomogeneous strain features, we can assume that $\Gamma_s$ in equation (\ref{eqn:g_broad}) is zero and that the Gaussian broadening in Fig. \ref{fig:peak} (b) is only due to instrumental contributions. Therefore, fitting the Gaussian widths with the Caglioti function will produce a triplet of U, V and W values which describe the IRF. It should be noted that the result of the fitting is affected by the choice of the units expressing the angle. In this work, the Gaussian FWHMs were expressed in degrees. Furthermore, the Williamson-Hall plot in Fig. \ref{fig:peak} (a), based on equation (\ref{eqn:williamson-hall}), provides the X and Y values which are characteristics of the x-ray beam bandwidth and the specimen grain size, respectively. All these fitting parameters are reported in Table \ref{tab:irf} with their respective standard errors.

\begin{table}
\centering
\caption{Fitting parameters of the Caglioti function (U, V, W) and of the Williamson-Hall plot (X, Y), with their respective standard errors.}
\begin{tabular}{ | c | c | c |}
\hline
 & Value & Standard error \\
\hline\hline

U &  \num{2.6912e-3}  &  \num{7.1727e-4}  \\
V &  \num{1.2460e-3}  &  \num{4.8220e-4}  \\
W &  \num{5.2366e-05}  &  \num{7.3779e-05}  \\
\hline
X &	\num{7.7621e-4}  &   \num{2.4310e-5}	\\
Y &	\num{1.4088e-4} & \num{8.3376e-06}	\\
\hline

\end{tabular}

\label{tab:irf}
\end{table}

\subsection{Refraction at grazing-incidence} \label{sec:refraction}

\begin{figure}
\includegraphics{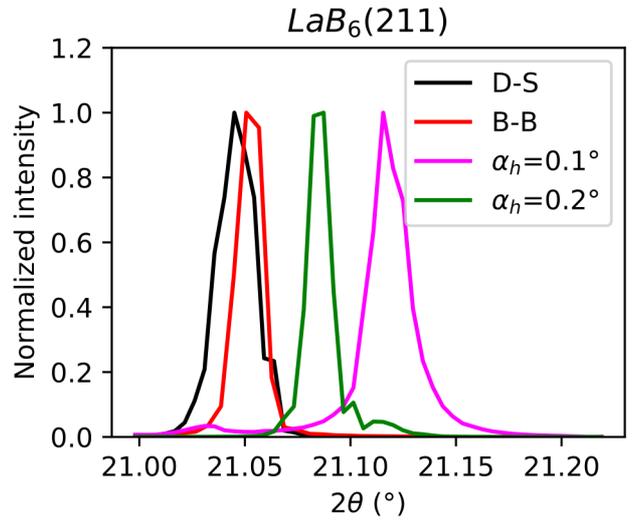}[ht!]
\caption{\ce{LaB6}(211) peak for different experimental geometries: Debye-Scherrer, Bragg-Brentano, GIXRD out-of-plane at $\alpha$=0.1° and 0.2°. The shift decreases with an increasing incidence angle and can be explained by refraction inside the \ce{LaB6} film.}
\label{fig:refr}
\end{figure}

\par A sample of \ce{LaB6} spin-coated on a Si(100) substrate was measured in Bragg-Brentano and grazing-incidence geometries, using a beam energy of 20 keV. The sample was prepared by mixing 500 mg of a \ce{LaB6} powder (Sigma-Aldrich, nominal grain size = 10 $\mu m$) with 10 mg of ethyl cellulose (Sigma-Aldrich, 48.0-49.5 \% (w/w) ethoxyl basis) as a binding agent, in 2 mL of ethanol. The mixture was spin-coated on a 6 mm square of Si(100) at 800 rpm. Fig. \ref{fig:refr} shows the \ce{LaB6}(211) peak a for the different geometries employed in this work, namely Debye-Scherrer (taken from the capillary data in \S \ref{sec:rietveld}), Bragg-Brentano and GIXRD at fixed grazing-incidence of 0.1° and 0.2°. The plots exhibit a shift in the peak position from the theoretical diffraction angle, $2\theta_{calc}$, which can be explained by refraction of light in the \ce{LaB6} film covering the Si substrate. For small incidence angles and a thin film with material constant $\delta$, \citet{lim1987} modelled this shift as follows:

\begin{equation}
\Delta 2\theta = \frac{\delta}{\sin{2\theta_{calc}}} \times \left(
				2 + \frac{\sin{\alpha}}{\sin{2\theta_{calc}}} + \frac{\sin{2\theta_{calc}}}{\sin{\alpha}}\right).
\label{eqn:refr}
\end{equation}

\par The positions of the peaks in Fig. \ref{fig:refr} for $\alpha_h$=0.1° and 0.2° were calculated in terms of center of mass and used to calculate the experimental peak shift by subtracting  $2\theta_{calc}$=21.0467° for \ce{LaB6}(211). The experimental shift is reported against the theoretical shift calculated using equation (\ref{eqn:refr}) in Table \ref{tab:refr}. The discrepancy between the two set of values can be explained by density inhomogeneities in the \ce{LaB6} film, which also contains ethyl cellulose for the practical purpose of consolidating the film. Furthermore, the \ce{LaB6} used for this layered sample is not a standard and it could have a slightly varied lattice parameter from what reported in literature after the dissolution in ethanol. Although this kind of sample is not recommended for calibration of XRD setups, it could be used as a reference sample to determine instrumental broadening at grazing-incidence. In order to calibrate the detector distance and the motor positions accurately, a well-calibrated standard is always recommended.

Equation (\ref{eqn:refr}) provides a way of correcting the data which is especially significant in strain analysis. In fact, homogeneous strain $\epsilon$ is calculated as the relative deviation of the atomic spacing d in the strained material from the atomic spacing $d_0$ in the ideally un-strained material: 
\begin{equation}
\epsilon = \frac{d-d_0}{d_0}.
\label{eqn:h_strain}
\end{equation}

This means that a polycrystalline strained material measured in grazing-incidence geometry will diffract at $2\theta$ angles that slightly deviate from the values predicted by Bragg’s law not only because of refraction, but also because of strain. If the strain of such material needs to be determined, equation (\ref{eqn:refr}) helps to correct for refraction peak shifts.

\begin{table}
\centering
\caption{Experimental and theoretical $\Delta 2\theta$ due to refraction in a \ce{LaB6} film.}
\begin{center}
\begin{tabular}{ | c | c | c |}
\hline
Incidence angle & $\alpha$=0.2° & $\alpha$=0.1° \\

\hline \hline
Experimental shift (°)& 0.03841 & 0.07142 \\

Theoretical shift (°) & 0.03488  & 0.06909 \\

Relative error (\%) & 9.19  & 3.27 \\
\hline
\end{tabular}
\end{center}
\label{tab:refr}
\end{table}

\section{Conclusions}	\label{sec:conclusions}

\par The angle calculations presented in this work describe how to convert data measured by a (2+3) diffractometer equipped with an area detector into a 1-dimensional XRD pattern, more familiar to the chemistry and material science communities. We also collected some dispersed knowledge on the phenomena contributing to the peak widths and on the intensity corrections. Since the calculations and the corrections are energy-independent, they can be applied to data collected at beamlines operating with hard x-rays (8-30 keV) as well as with high energies (40-150 keV). The Python code used in this work for such corrections and angle calculations is available on Github (\url{github.com/giuseppe-abbondanza/pyLjus}).
\par Quantitative PXRD using surface diffractometers is not often done. This work should however facilitate quantitative studies using such instruments, including for example \textit{in situ} studies under grazing incidence conditions.
\par Furthermore, the calculations presented in this work can be modified to apply to diffractometers with other geometries (e.g., (2+2), 4-circle and 6-circle) and they can even apply to setups where the detector is not mounted on a motor and therefore has a fixed position, facing the direct beam. In this configuration, one can assume that effectively $\gamma$=$\delta$=0°.

\begin{center}
\textbf{Appendix}
\end{center}

\par Here we present the coordinate transformations to a assign a diffraction angle $2\theta$ and an azimuthal angle $\chi$ to every pixel of an area detector, given the distance $R$ of the detector from the center of the diffractometer and the nominal detector angles $\gamma$ and $\delta$. The angles and the wavevectors involved in the calculations are schematically shown in Fig. \ref{fig:vectors}. The following derivation is based on the reciprocal space coordinate calculations given by \citet{schleputz2010}. 
\par These calculations are valid for both horizontal geometry and vertical geometry. In both cases, the laboratory frame of reference has its zero-coordinates on the diffractometer center and the y-axis points in the direction of the synchrotron beam. Therefore, the wavevector of the incoming x-ray beam is

\begin{equation}
\label{eqn:kin}
\vec{k} = k
\begin{pmatrix}
0 \\
1 \\
0\\
\end{pmatrix},
\end{equation}

where $k = 2\pi/\lambda$ is the magnitude of the wavevector and $\lambda$ is the wavelength of the x-ray beam. Similarly, a generic scattered wavevector such as the one in Fig. \ref{fig:vectors} (a) can be represented as

\begin{equation}
\label{eqn:kout}
\vec{k'} = k 
\begin{pmatrix}
\cos{\delta} \sin{\gamma} \\
\cos{\delta} \cos{\gamma} \\
\sin{\delta} \\
\end{pmatrix}.
\end{equation}

In the assumption that the scattering is elastic (i.e., $\mid \vec{k} \mid = \mid \vec{k'} \mid = k$), the angle $2\theta$ can be found using the definition of scalar product as follows:

\begin{equation}
\label{eqn:dotp_def}
\vec{k} \cdot \vec{k'} = \mid \vec{k} \mid \mid \vec{k'} \mid \cos{2\theta},
\end{equation}

\begin{equation}
\label{eqn:dotp}
\vec{k} \cdot \vec{k'} = k^2 \cos{\delta} \cos{\gamma},
\end{equation}

\begin{equation}
\label{eqn:2th}
2\theta = \arccos(\cos{\delta} \cos{\gamma}).
\end{equation}

The azimuthal $\chi$ angle can be calculated as 

\begin{equation}
\chi = \arctan\left(\frac{\tan{\delta}}{\tan{\gamma}}\right).
\label{eqn:chi}
\end{equation}

Therefore, when $\delta$ and $\gamma$ are known for each pixel of the detector, it is possible to assign a $2\theta$ and a $\chi$ value to each pixel. To this aim, we need to calculate the exact position of the detector center.


\par We define the detector center as the point that is impinged by the direct beam when the detector is in the zero position (i.e., when $\gamma = \delta = \nu = 0$, as shown in figure \ref{fig:diff}). In this position, the coordinates of the detector center are given by calculating the \enquote{center of mass} of an image recorded while the direct beam impinges the detector:

\begin{equation}
\label{eqn:mcen}
\vec{c} = \frac{1}{S} \sum_{i,j}I(i,j)\vec{p}(i,j),
\end{equation}

where $(i,j)$ are the detector pixel coordinates, $\textbf{c}$ is the vector pointing to the mass center, $S$ is the sum of all intensities detected in the image, $I(i,j)$ and $\textbf{p}(i,j)$ are the intensity and the vector describing the position of the $(i,j)$ pixel.

\par Let us denote the coordinates of the detector center found in equation (\ref{eqn:mcen}) with $(c_x, c_z)$. A pixel with coordinates $(i,j)$ has an offset $(\Delta x, \Delta z)$ from the detector center given by:

\begin{equation}
\label{eqn:deltax}
\Delta x = (c_x-i)w_x,
\end{equation}

\begin{equation}
\label{eqn:deltaz}
\Delta z = (c_z-j)w_z,
\end{equation}

where $w_x$ and $w_z$ are the width of the pixel along the x and z directions, respectively (for a Pilatus 100K detector $w_x = w_z = 172$ $\mu m$). Fig. \ref{fig:vectors} (b) illustrates the typical offsets in pixel position described above. Here we assign the (0,0) coordinates to the upper left detector pixel.
When the detector is at the zero position, the position of a generic $(i,j)$ pixel in the laboratory coordinates is simply

\begin{equation}
\label{eqn:zero_angle}
\begin{pmatrix}
x_p \\
y_p \\
z_p \\
\end{pmatrix}
=
\begin{pmatrix}
\Delta x \\
R \\
\Delta z \\
\end{pmatrix}.
\end{equation}

For non-zero detector angles, the new $(x_p, y_p, z_p)$ coordinates of the $(i,j)$ pixel are found by rotating the vector $(\Delta x, R, \Delta z)$ by $\gamma$, $\nu$ and $\delta$ around the z, y and x axes, respectively. Therefore, the pixel position in the laboratory frame of reference becomes

\begin{equation}
\label{eqn:nonzero_angle}
\begin{pmatrix}
x_p \\
y_p \\
z_p \\
\end{pmatrix}
=
\vec{\Gamma} \vec{\Delta} \vec{N}
\begin{pmatrix}
\Delta x \\
R \\
\Delta z \\
\end{pmatrix},
\end{equation}

where $\vec{\Gamma}$, $\vec{\Delta}$ and $\vec{N}$ are the matrices describing the rotations around the z, x and y axes, defined as follows:

\begin{equation}
\label{eqn:GAM_h}
\vec{\Gamma} = \vec{R_z}(\gamma) = 
\begin{pmatrix}
\cos{\gamma} & -\sin{\gamma} & 0 \\
\sin{\gamma} & \cos{\gamma} & 0 \\
0 & 0 & 1 \\
\end{pmatrix},
\end{equation}

\begin{equation}
\label{eqn:DEL_h}
\vec{\Delta} = \vec{R_x}(\delta) = 
\begin{pmatrix}
1 & 0 & 0 \\
0 & \cos{\delta} & -\sin{\delta} \\
0 & \sin{\delta} & \cos{\delta} \\
\end{pmatrix}.
\end{equation}

\begin{equation}
\label{eqn:N_h}
\vec{N} = \vec{R_y}(\nu) = 
\begin{pmatrix}
\cos{\nu} & 0 & \sin{\nu} \\
0 & 1 & 0 \\
-\sin{\nu} & 0 & \cos{\nu} \\
\end{pmatrix},
\end{equation}

The order of matrix multiplication depends on how the diffractometer is built up. Since the $\gamma$-circle holds the $\delta$-circle, which in turn sustains the $\nu$-motor, the order of matrix multiplication is $\vec{R_z}$, $\vec{R_x}$ and $\vec{R_y}$.
\par The angle values for a generic pixel $\gamma_p$ and $\delta_p$ can be calculated as follows:

\begin{equation}
\gamma_p = \arctan\left(\frac{z_p}{y_p}\right),
\label{eqn:gamma_p}
\end{equation}

\begin{equation}
\delta_p = \arcsin\left(\frac{x_p}{d}\right),
\label{eqn:delta_p}
\end{equation}

where d is the distance of the pixel from the center of the diffractometer given by

\begin{equation}
d = \sqrt{{\Delta x}^2 + R^2 + {\Delta z}^2}.
\label{eqn:dist}
\end{equation}

\begin{center}
\textit{Detector slits}
\end{center}

\par
\begin{figure}[ht!]
\includegraphics{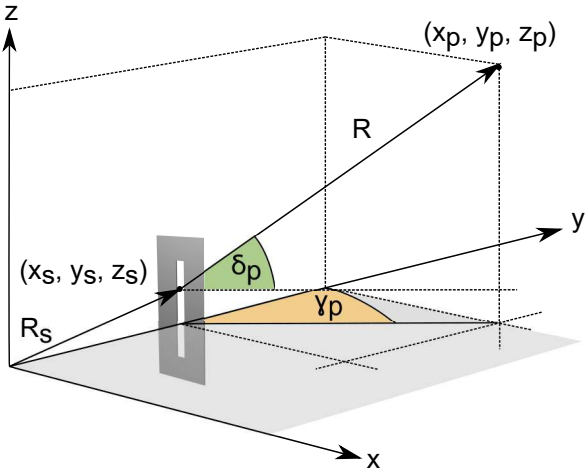}
\caption{Diagram showing how the detector angles are affected by the use of guard slits.}
\label{fig:slits}
\end{figure}

 Apertures and slits are used in PXRD to control the beam size, divergence and angular resolution. High-density metals are often used in the fabrication of slits due to their low x-ray transmission. 
When the detector slits are engaged, part of the diffracted x-rays is blocked and the effect of the beam footprint on the peak broadening is reduced (see \S \ref{sec:footprint}). The detected signal can be seen as originating from a point within the slit aperture rather than from the diffractometer centre. Therefore, we can derive a new set of \textit{effective} detector angles ($\gamma_p$, $\delta_p$), as illustrated in Fig. \ref{fig:slits}.

The $\nu$ rotation axis is always perpendicular to the aperture plane for every nominal detector angle ($\gamma$, $\delta$), i.e. the slit aperture rotates with the detector. In this assumption, the coordinates of the aperture in the laboratory coordinate system are given by

\begin{equation}
\begin{pmatrix}
x_s \\
y_s \\
z_s \\
\end{pmatrix}
= \vec{\Gamma} \vec{\Delta} 
\begin{pmatrix}
0 \\
R_s \\
0 \\
\end{pmatrix},
\label{eqn:slit_coord}
\end{equation}

where $R_s$ is the slit distance from the diffractometer center. The effective detector angles are now dependent on the distance between the pixel and the slit and are given by

\begin{equation}
d_{d,s} = \sqrt{(x_d - x_s)^2 + (y_d - y_s)^2 + (z_d - z_s)^2},
\label{eqn:d_slit}
\end{equation}

\begin{equation}
\gamma_p = \arctan\left(\frac{z_p - z_s}{y_p - y_s}\right),
\label{eqn:gammap_slit}
\end{equation}

\begin{equation}
\delta_p = \arcsin\left(\frac{x_p - x_s}{d_{d,s}}\right).
\label{eqn:deltap_slit}
\end{equation}

\begin{center}
\textit{Vertical geometry}
\end{center}

\par In the vertical geometry the surface normal of the sample and the y-axis lay in the horizontal plane, while the x-axis points upwards. The equations for the incoming beam wavevector and those for the offset from the detector center are the same as in the horizontal geometry, i.e., equations (\ref{eqn:kin}), (\ref{eqn:deltax}) and (\ref{eqn:deltaz}), while the generic scattering vector is given by 

\begin{equation}
\label{eqn:kout_vert}
\vec{k'} = k 
\begin{pmatrix}
\sin{\delta} \\
\cos{\delta} \cos{\gamma} \\
\cos{\delta} \sin{\gamma} \\
\end{pmatrix}.
\end{equation}

Since the y-component has not changed, the equation for the $2\theta$ angle is the same as in equation (\ref{eqn:2th}). The equation for the $\chi$  angle (\ref{eqn:chi}) is also the same in both geometries. The rotation matrices, however, are different in this frame of reference and are given by

\begin{equation}
\label{eqn:GAM_v}
\vec{\Gamma} = \vec{R_x}(\gamma) = 
\begin{pmatrix}
1 & 0 & 0 \\
0 & \cos{\gamma} & -\sin{\gamma} \\
0 & \sin{\gamma} & \cos{\gamma} \\
\end{pmatrix}
\end{equation}

\begin{equation}
\label{eqn:N_v}
\vec{N} = \vec{R_y}(\nu) = 
\begin{pmatrix}
\cos{\nu} & 0 & \sin{\nu} \\
0 & 1 & 0 \\
-\sin{\nu} & 0 & \cos{\nu} \\
\end{pmatrix}
\end{equation}

\begin{equation}
\label{eqn:DEL_v}
\vec{\Delta} = \vec{R_z}(\delta) = 
\begin{pmatrix}
\cos{\delta} & \sin{\delta} & 0 \\
-\sin{\delta} & \cos{\delta} & 0 \\
0 & 0 & 1 \\
\end{pmatrix}
\end{equation}

As in the case of horizontal geometry, equation (\ref{eqn:nonzero_angle}) applies to the calculation of pixel coordinates.

\begin{center}
\textbf{Acknowledgements}
\end{center}

Measurements were performed on the I07 beamline at the Diamond Light Source. This work was financially supported by the Swedish Research Council through the Röntgen-Ångström Cluster “In-situ High Energy X-ray Diffraction from Electrochemical Interfaces (HEXCHEM)” (Project no. 2015-06092) and project grant "Understanding and Functionalization of Nano Porous Anodic Oxides" (project no. 2018-03434) by the Swedish Research Council. G. Abbondanza acknowledges financial support from NanoLund under grant p05-2017. We wish to thank R. Felici for valuable discussions.

\printbibliography

\end{document}